# Optofluidic ptychography on a chip

Pengming Song,‡a Chengfei Guo,‡*a Shaowei Jiang,‡a Tianbo Wang,a Patrick Hu,b Derek Hu,c Zibang Zhang,a Bin Feng,a and Guoan Zheng*a

**We report the implementation of a fully on-chip, lensless microscopy technique termed optofluidic ptychography. This imaging modality complements the miniaturization provided by microfluidics and allows the integration of ptychographic microscopy into various lab-on-a-chip devices. In our prototype, we place a microfluidic channel on the top surface of a coverslip and coat the bottom surface with a scattering layer. The channel and the coated coverslip substrate are then placed on top of an image sensor for diffraction data acquisition. Similar to the operation of flow cytometer, the device utilizes microfluidic flow to deliver specimens across the channel. The diffracted light from the flowing objects is modulated by the scattering layer and recorded by the image sensor for ptychographic reconstruction, where high-resolution quantitative complex images are recovered from the diffraction measurements. By using an image sensor with a 1.85 μm pixel size, our device can resolve the 550 nm linewidth on the resolution target. We validate the device by imaging different types of biospecimens, including *C. elegans*, yeast cells, *paramecium*, and *closterium sp.* We also demonstrate high-resolution ptychographic reconstruction at a video framerate of 30 frames per second. The reported technique can address a wide range of biomedical needs and engenders new ptychographic imaging innovations in a flow cytometer configuration.**

## Introduction

Optofluidic platforms bring microfluidics and optics together to create highly versatile systems that can be incorporated into lab-on-a-chip devices[1-7]. Combining fluids and light has

a. *Department of Biomedical Engineering, University of Connecticut, Storrs, CT, 06269, USA.*
b. *Department of Computer Science, University of California Irvine, Irvine, CA, 92697, USA.*
c. *Amador Valley High School, Pleasanton, CA, 94566, USA*
*Email: chengfei.guo@uconn.edu or guoan.zheng@uconn.edu*
† Electronic Supplementary Information (ESI) available: Figs. S1-S5, Notes 1-2, Movie S1. See DOI: 10.1039/x0xx00000x
‡ These authors contributed equally to this work.

produced various creative devices in the past decade, including optofluidic laser[8], configurable optofluidic prism[9], tunable optofluidic switch[10], lensless optofluidic microscopy[11-15], among others. In parallel with the development of optofluidic technologies, ptychography has emerged as an enabling coherent diffraction imaging (CDI) approach for X-ray and electron microscopy. The original concept was developed to address the phase problem of crystallography[16]. Its modern form was brought to fruition by adopting the framework of iterative phase retrieval[17]. In a typical implementation[17-23], the specimen is laterally translated through a spatially confined illumination beam and the diffraction patterns are recorded at the reciprocal space. The reconstruction process iteratively imposes two sets of constraints. The confined illumination beam limits the physical extent of the object for each measurement and serves as the support constraint in the real space. The diffraction measurements serve as the Fourier magnitude constraints in the reciprocal space. In the visible region, it is possible to swap the real space and the reciprocal space using a lens. The development of Fourier ptychography is an example in this direction[24, 25]. In the past decade, ptychography has become an indispensable imaging tool in most synchrotron and national laboratories worldwide[26, 27].

Here we report an integrated ptychographic imaging technique termed optofluidic ptychography. This new imaging modality complements the miniaturization provided by microfluidics and allows the integration of ptychographic microscopy into various lab-on-a-chip devices. To the best of our knowledge, it is the first fully on-chip implementation of ptychography by integrating the concepts of flow cytometry and ptychographic CDI. To facilitate this development, we report a procedure to accurately track the positional shift of the objects passing through the microfluidic channel, thereby eliminating the requirement for precise positioning in traditional ptychographic implementations. By using an image sensor with a 1.85 nm pixel size, our device can resolve the 550 nm linewidth on the resolution target and recover the object phase in a quantitative manner. We validate our device by imaging different types of biospecimens, including *C. elegans*, yeast cells, *paramecium*, and *closterium sp.* We also develop a procedure to recover high-resolution video of objects passing through the channel, where the interaction between the fluid and the object can be monitored in high resolution. The combination of ptychography and microfluidic techniques in a flow cytometer configuration is a significant step towards a cost-effective, chip-scale imaging solution that addresses the needs of biomedical and biological sciences.

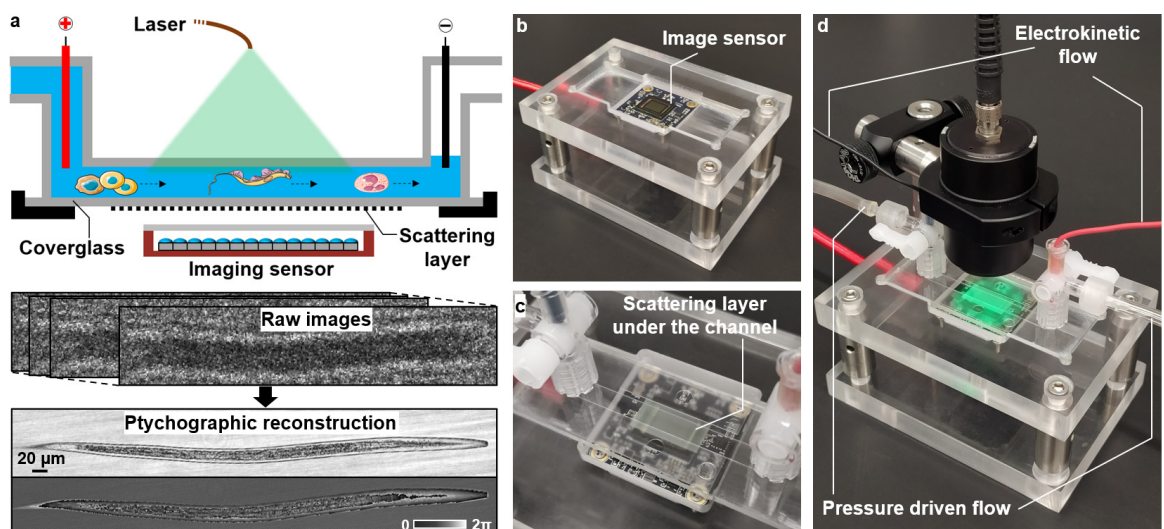

**Fig. 1** Operating principle of optofluidic ptychography. (a) A microfluidic channel is placed on the top surface of a coverslip. The bottom surface is coated with a scattering layer. The channel and the coated coverslip substrate are then placed on top of an image sensor for diffraction data acquisition. Similar to the operation of flow cytometer, the device utilizes microfluidic flow to deliver specimens across the channel. The diffracted light from the flowing objects is modulated by the scattering layer and recorded by the image sensor for ptychographic reconstruction. (b) The image sensor mounted on an acrylic case. (c) The microfluidic channel and the coverslip substrate with scattering layer coating. (d) The prototype device where we translate the specimens by either electrokinetic flow or pressure-driven flow.

## Methods

### Optofluidic ptychography

Figure 1 shows the operation principle and the prototype device of the optofluidic ptychography technique. As shown in Fig. 1a, we place a microfluidic channel on the top surface of a coverslip and coat the bottom surface with a scattering layer. The channel and the coated coverslip substrate are then placed on top of an image sensor for diffraction data acquisition. Similar to the operation of flow cytometer, the device utilizes microfluidic flow to deliver specimens across the channel. The diffracted light from the flowing objects is modulated by the scattering layer and recorded by the image sensor. The scattering layer serves as a computational lens[28-32] for the ptychographic imaging process[33-36], where high-resolution quantitative complex object images are recovered from the diffraction intensity measurements. The translation of the specimen by microfluidic flow allows the object exit wave to be probed by different parts of the scattering layer, thereby providing phase diversity measurements[37] and enabling sub-pixel resolving capability[12, 13, 35]. The bottom panel of Fig. 1a shows the captured raw images and the corresponding ptychographic reconstructions of intensity and phase.

Our prototype device is shown in Fig. 1b-1d, where we use a monochrome image sensor with a 1.85 μm pixel size for diffraction data acquisition (Sony IMX 226). The images are acquired at 30 frames per second as the objects flow across the channel. Figure 1c shows the microfluidic channel and the coverslip substrate. The channel width is 5 mm and the channel height is 0.5 mm. On the bottom surface of the coverslip substrate, we coat a thin layer of polystyrene beads as the scattering lens (Fig. 1c). The distance between the scattering layer and the sensing surface of the image sensor is ~1 mm. We also note that the microfluidic channel does not need to be tilted at the x-y plane for pixel super resolution imaging as demonstrated in the lensless optofluidic microscope[12, 13, 15]. Instead, we bypass the resolution limit defined by the pixel size via the scattering layer and the sub-sampling model in ptychographic reconstruction[38]. Figure 1d shows the entire prototype device in operation, where we couple a 532 nm diode laser to a single-mode fiber for sample illumination. The measured intensity at the detector plane is ~15 mW. Compared to LED illumination, the use of laser beam allows us to shorten the exposure time to <0.2 ms, addressing the motion blur issue in the image acquisition process.

### Electrokinetic flow and pressure-driven flow

As shown in Fig. 1d, we employ two techniques to drive the objects across the channel: electrokinetic flow and pressure-driven flow[11]. For electrokinetic flow, we apply a DC voltage of ~36 V between the inlet and outlet in Fig. 1d, and the flow speed is 40-80 μm/s. Electrokinetic flow can be used for relatively small size objects such as cells or cell clusters. It can minimize sample rotation during the translation process. For pressure-driven flow, we simply inject the sample with liquid in the inlet and create a pressure difference between the two ends of the microfluidic channel. The flow speed is 100-400 µm/s. An object flowing in the channel will receive a torque from the parabolic velocity profile caused by the nonslip boundary condition on the channel sidewalls. For both electrokinetic flow and pressure-driven flow, we typically acquire 90 raw images in ~3 seconds for ptychographic reconstruction. As we will demonstrate later, both the transverse translational shift and the rotation angle need to be inferred from the raw diffraction measurements.

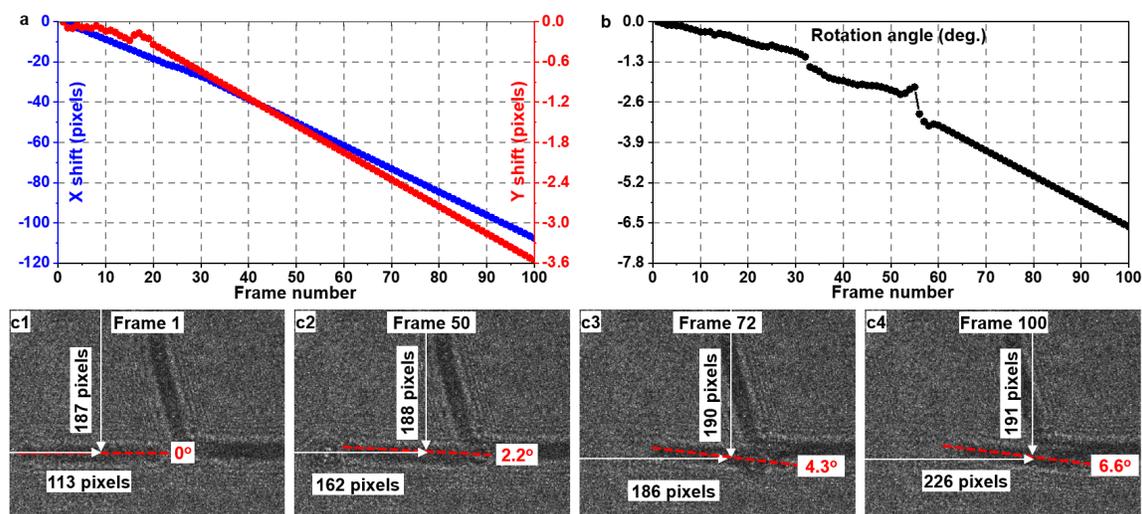

**Fig. 2** Tracking of the positional shift and the rotation angle of objects passing through the channel. (a) The tracked positional shifts of a *C. elegan* passing through the channel via pressure-driven flow. (b) The rotation angle of the *C. elegan*. (c) The raw images are processed to generate digital in-line holographic reconstruction for the tracking process (also refer to Fig. S1). The background of the raw images shows two *C. elegan* sticked to the bottom of the channel, causing the flowing *C. elegan* to rotate.

### Tracking the flowing objects

In a regular ptychographic implementation, the positional shift can be directly obtained from the mechanical stages. In our device, both electrokinetic flow and pressure-driven flow induce non-uniform positional shift and sample rotation. The scattering layer of our device further modulates the diffracted light from the objects. Thus, it is challenging to directly recover the positional shift and rotation via correlation analysis[12, 33].

In our implementation, we first average all measurements and get an initial estimate of the scattering layer. The captured raw images are divided by this estimate to remove the modulation pattern. We then digitally propagate the processed raw images to the plane of the microfluidic channel; the phase is assumed to be 0 as an approximation. This process is the same as that in digital in-line holography[39]. Figure S1 shows the difference between the raw image (Fig. S1a), the raw image divided by the estimate of the scattering layer (Fig. S1b), the digital in-line holographic reconstruction (Fig. S1c), and the final super-resolution ptychographic reconstruction (Fig. S1e). Based on the digital in-line holographic reconstruction, the positional shift and the rotation angle can be recovered using the following optimization procedure:

$$(\hat{x}_i, \hat{y}_i, \hat{\theta}_i) = \underset{x_i, y_i, \theta_i}{\arg\min} \sum_{(x,y) \in C} [T_{x_i, y_i, \theta_i} \{O_1(x, y)\} - O_i(x, y)]^2, \quad (1)$$

where $O_i(x, y)$ is the digital in-line holographic reconstruction for the $i^{th}$ captured image $I_i(x, y)$. In our implementation, $O_i(x, y) = \left| sqrt(\frac{I_i(x,y)}{\sum_{j=1}^J I_j(x,y)/J}) * PSF_{free}(-d) \right|$, where '$J$' is the total number of the captured images. '*' denotes the convolution operation. We use $PSF_{free}(d)$ to model the point spread function (PSF) for free-space propagation over a distance $d$, and the negative sign means back-propagation to the plane of the microfluidic channel. The image summation in Eq. (1) is performed over the region $C$ of the tracked object and $T_{x_i, y_i, \theta_i}\{\cdot\}$ is a transform operation of an image with transverse positional shifts and rotation. Based on the two digital in-line holographic reconstructions $O_1(x, y)$ and $O_i(x, y)$, Eq. (1) aims to recover the relative transverse positional shift ($x_i, y_i$) and the rotation angle $\theta_i$ between them. The detailed implementation and the closed-form solution[40] of $(\hat{x}_i, \hat{y}_i, \hat{\theta}_i)$ in Eq. (1) can be found in Supplementary Note 1. We note that the transform $T_{x_i, y_i, \theta_i}\{\cdot\}$ discussed above assumes a rigid object. It is possible to further consider the object's internal distortion or bending using a more general transform.

Figure 2 shows a typical tracking result using the optimization process in Eq. (1). Figures 2a and 2b show recovered transverse positional shifts and the rotation angle of a *C. elegans* under the pressure-driven flow. Figure S2 shows the tracking results of yeast cells under the electrokinetic flow. The same procedure has also been applied to a resolution target in our characterization experiment in a later section.

### Ptychographic reconstruction

With the recovered $(\hat{x}_i, \hat{y}_i, \hat{\theta}_i)$, we can perform ptychographic reconstruction using the captured raw images. All previous ptychographic implementations only consider object translation in the acquisition process. One key consideration of the reported platform is that we need to properly model both the translation and rotation of the object. The forward imaging model of our optofluidic ptychography platform can be expressed as

$$I_i(x,y) = \left| \begin{array}{c} T_{x_i, y_i, \theta_i} \{ E(x,y) \} * PSF_{free}(d_1) \cdot D(x,y) \\ * PSF_{free}(d_2) * PSF_{pixel} \end{array} \right|^2_{\downarrow M}, \quad (2)$$

where $I_i(x,y)$ is the i$^{th}$ captured image, $E(x,y)$ is the exit wave of the complex object, $D(x,y)$ is the complex profile of the scattering layer, $d_1$ is the distance between the object exit wave and the scattering layer, $d_2$ is the distance between the scattering layer and the sensor surface, '$\cdot$' stands for point-wise multiplication, and we use $PSF_{pixel}$ to model the PSF of the pixel response. Due to the relatively large pixel size, the captured image is a down-sampled version of the diffraction pattern, and we use '$\downarrow M$' in the subscript of Eq. (2) to represent this down-sampling process. The ptychographic reconstruction process aims to recover $E(x,y)$ from measurements $I_i(x,y)$ ($i = 1,2,3...$).

As shown in Fig. S3, conventional image rotation via spatial interpolation is not a reversible operation. Therefore, it will accumulate the interpolation error as we perform image rotation many times during the iterative reconstruction process. In Supplementary Note 2, we discuss the detailed implementation of Fourier interpolation method[41] for image rotation that is reversible. We also note that object translation in ptychographic reconstruction is also implemented in the Fourier domain[42] with sub-pixel accuracy. Therefore, the employed image rotation operation is fully compatible with the current routine by adding shearing operations in the Fourier domain. The entire reconstruction process is similar to that in near-field Fourier ptychography[43], with the speckle patterned replaced by the profile of the scattering layer. The rotation operation can be added to the translation process in the Fourier domain. For different objects tracked over the field of view of the microfluidic channel, we can recover their corresponding translation and rotation for reconstruction, similar to that demonstrated in Ref. [44].

## Results

### Characterization of imaging performance

We first validate the imaging performance of the prototype device using both a resolution target and a quantitative phase target in Fig. 3. In equivalent to the object translation within the channel, we move the targets across the coverslip substrate and acquire the diffraction measurements for reconstruction. The same tracking process discussed in the previous section is employed.

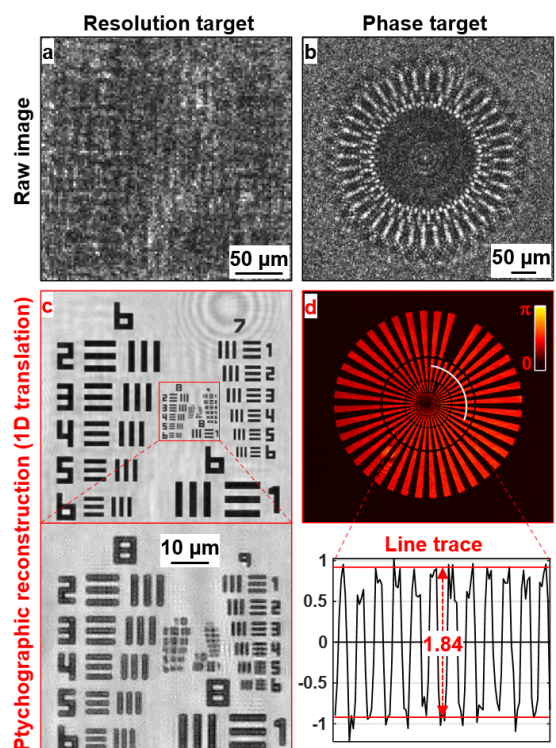

**Fig. 3** Validation using a resolution and a phase target. Raw images (a-b) and the reconstruction (c-d) of the two targets. We can resolve the 550 nm linewidth of the resolution target.

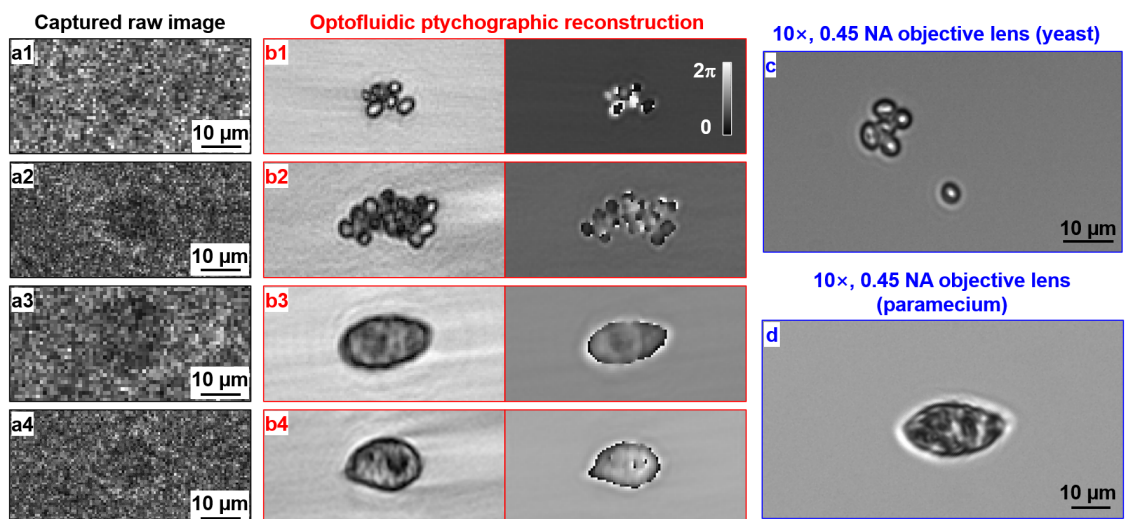

**Fig. 4** Optofluidic ptychographic reconstruction using electrokinetic flow. (a) Captured raw image for yeast cells and *paramecium* cells. (b) The recovered intensity and phase images based on 90 raw measurements and 3 iterations of the ptychographic process. (c-d) Images of the same species captured using a regular microscope with a 10×, 0.45 NA objective (not the same cells in b).

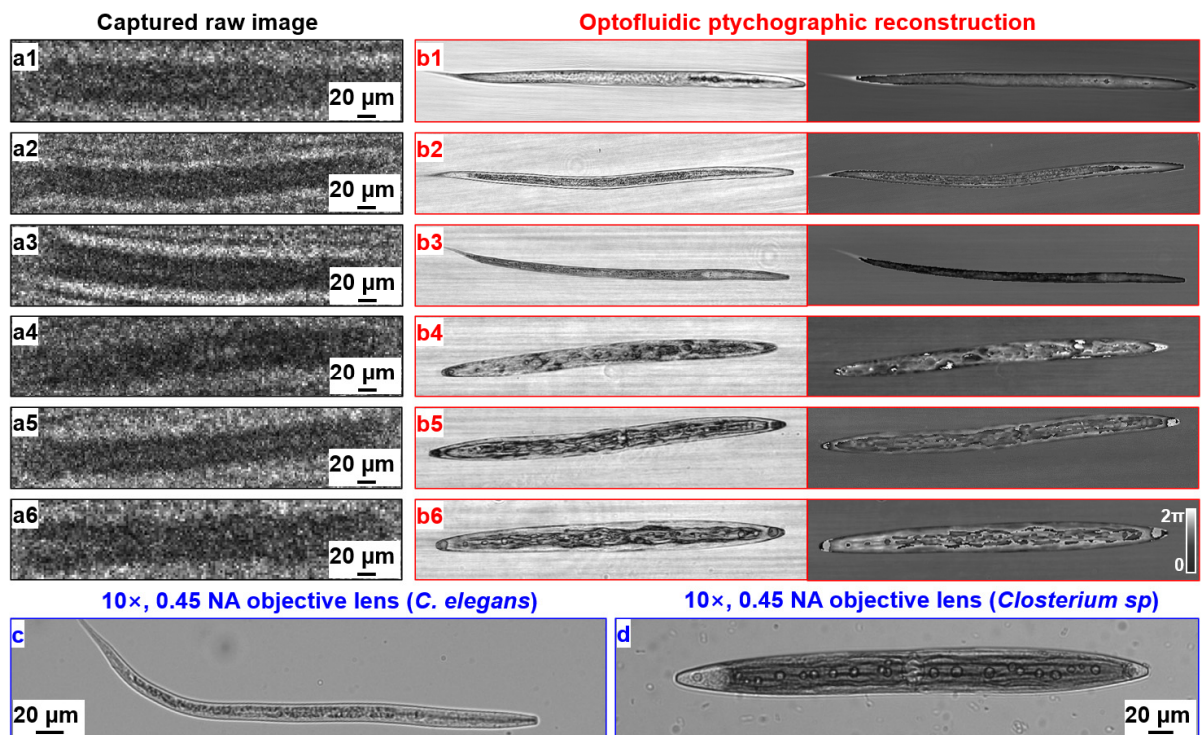

**Captured raw image**

**Optofluidic ptychographic reconstruction**

a1 · a2 · a3 · a4 · a5 · a6

b1 · b2 · b3 · b4 · b5 · b6

**10×, 0.45 NA objective lens (*C. elegans*)**

**10×, 0.45 NA objective lens (*Closterium sp*)**

c · d

**Fig. 5** Optofluidic ptychographic reconstruction using pressure-driven flow. (a) Captured raw images for *C. elegans* and *closterium sp*. (b) The recovered intensity and phase images based on 90 raw measurements and 3 iterations of the ptychographic process. (c-d) Images of the same species captured using a regular microscope with a 10×, 0.45 NA objective (not the same cells in b).

Figure 3a-3b show the captured raw images of the targets. Figure 3c-3d show the ptychographic reconstructions using 500 raw images and 5 iterations of the phase retrieval process. From Fig. 3c, we can resolve the 550 nm linewidth of group 9, element 6 of the target. In this demonstration, we obtain a 3.4-fold resolution gain using the image sensor with a 1.85 μm pixel size. For the recovered phase image, we also compare it to the ground truth phase in Fig. 3d, where they are in good agreement with each other. This experiment validates the super-resolution performance and the quantitative nature of the reported platform. Figure S4 shows the recovered complex profile of the scattering layer from a calibration experiment. Figure S5 further shows the recovered images of the resolution target with different numbers of raw measurements.

**Imaging with electrokinetic and pressure-driven flows**

Figure 4 shows the imaging performance using the electro-kinetic flow. The captured raw images of yeast cells and *paramecium* cells are shown in Fig. 4a, and their ptychographic reconstructions are shown in Fig. 4b. In Fig. 4c, we show the images of the same species captured using a regular microscope platform with a 10×, 0.45 NA objective lens. We can see that the ptychographic reconstruction has a similar image quality as that of the regular microscope platform. In Fig. 5, we show the imaging performance of our device using the pressure-driven flow. The captured raw images of *C. elegans* and *closterium sp* are shown in Fig. 5a, and their ptychographic reconstructions are shown in Fig. 5b. In Fig. 5c, we also show the images of the same species captured using a regular microscope platform with a 10×, 0.45 NA objective lens.

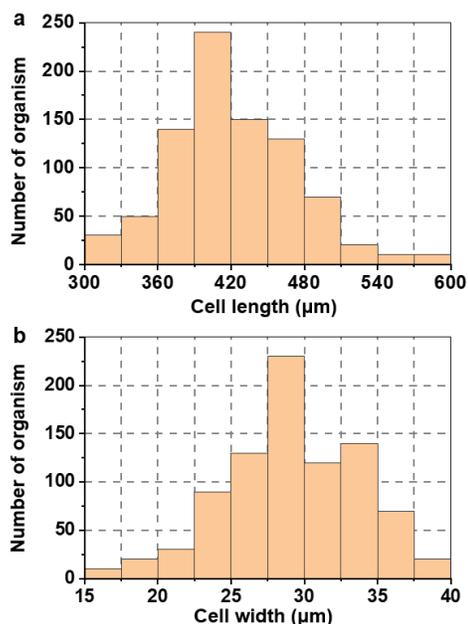

**Fig. 6** Cytometric analysis of cell length and width from the optofluidic ptychographic reconstructions.

In the pressure-driven flow experiment, we also perform a cytometric analysis of the objects passing through the channel.

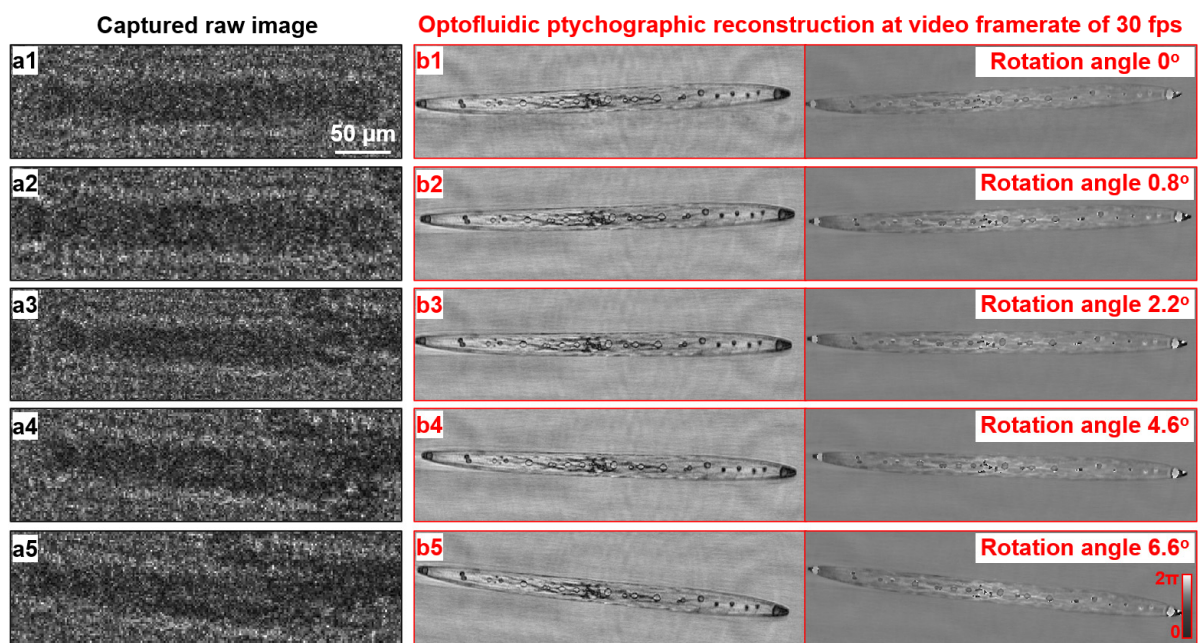

**Fig. 7** Sequential optofluidic ptychographic reconstruction of rotating objects at a video framerate of 30 fps. (a) Captured raw images for the *closterium sp* cell. (b) The recovered intensity and phase images at different time points as the object rotates to different angles. Also refer to Movie S1.

With a ~5-min dataset, we have identified ~1000 *closterium sp* in our experiment and Fig. 6 shows the measured length and width of the cells. The integration of ptychography with flow cytometry may open up new possibilities for high-throughput cell analysis and screening.

**Ptychographic reconstruction at 30 frames per second**

The optofluidic ptychography platform can also recover high-resolution videos of the objects as they flow across the channel and interact with the fluid. This can be accomplished using a sliding window sampling approach[12] in combination with the optofluidic ptychographic reconstruction. In Fig. 7 and Movie S1, we demonstrate the reconstruction of a high-resolution video of *closterium sp* flowing across the channel. To reconstruct the video, we choose the first 100 raw images and recover the first high-resolution object image with 3 iterations. This reconstruction is then used as an initial guess for the next reconstruction with the 101st frame. The process is repeated for all subsequent frames. The reconstructions shown in Fig. 7 are obtained from the sequence of the reconstruction.

Super-resolution optofluidic ptychographic reconstruction at the camera framerate may find many practical applications, including monitoring the dynamic behaviour of a sample, tracking rotation and deformation of cells, and creation of synthetic 'video zoom' where a region of interest within the video sequence can be enlarged and enhanced[12].

## Discussion and conclusion

In summary, we have demonstrated a fully on-chip, lensless microscopy imaging technique termed optofluidic ptychography. This new imaging modality complements the miniaturization provided by microfluidics and allows the integration of high-resolution ptychographic microscopy into various lab-on-a-chip devices. To the best of our knowledge, it is the first demonstration of ptychography in a flow cytometer configuration. We develop a procedure for tracking the object motion within the microfluidic channel. This procedure can be useful for other ptychographic implementations where the object motion needs to be precisely recovered for the phase retrieval process. We also demonstrate the use of a sliding window strategy to perform high-resolution ptychographic reconstruction at 30 frames per second. This strategy has the potentials to monitor dynamic behaviours of cells and slow-moving objects in high resolution.

We note that, in a regular ptychographic implementation, a spatially confined probe beam is used for object illumination. The diffraction patterns are acquired as the object translates across the beam. This scheme can be directly implemented in a flow cytometer configuration where the flowing objects are illuminated by a spatially confined laser beam. An image sensor can be used to record the corresponding diffraction patterns as the object flows across the channel. This scheme, however, has two major drawbacks. First, the confined probe beam limits the imaging throughput as we can only track one or few objects at a time. Second, it is challenging to infer the positional shifts and the rotation angles of objects that pass through the confined probe beam.

In this work, we employ full-field illumination that covers the entire microfluidic channel, addressing the throughput limit of the traditional ptychographic implementation. The ptychographic probe is provided by the scattering layer coated on the coverslip substrate. By using an image sensor with a 1.85 μm pixel size, our device can resolve the 550 nm linewidth on the resolution target. The use of laser in our implementation allows us to capture images with a short exposure time (<0.2 ms), addressing the motion blur issue of the flowing objects. The image quality can be further improved by reducing the spatial coherence of the laser source.

In our imaging model, a transform $T_{x_i, y_i, \theta_i}\{\cdot\}$ is applied to the object for modelling translation and rotation. This transform assumes the object to be rigid; no internal distortion or bending is allowed. Our ongoing effort is to adopt a more general transform that can model the internal distortion of the object. Our goal is to continuously monitor the behaviour of microorganisms with internal distortion at the camera framerate. We also not that, the dominant force effect in the fluid domain can be estimated according to the object position. It is possible to recover this position-based deformation information using the reported platform. If successful, it will open up new opportunities for quantitative cell screening and other fields.

Lastly, the current platform cannot be used for high-resolution fluorescence imaging. One solution is to apply speckle illumination for fluorescence imaging and jointly recover the speckle pattern and the high-resolution object as demonstrated in Ref.[45].

The development of ptychographic imaging in a flow cytometer configuration, currently in its infancy, will continue to improve in performance and expand in applications. The reported optofluidic ptychography technique is among the first steps in this direction.

## Author contributions

P. S., C. G., and S. I. contributed equally to the work and develop the reported prototype platform. T.B., P.H., and D.H. performed image reconstruction and data analysis. G. Z. conceived the project and designed the experiment. All authors contributed to writing and correcting the manuscript.

## Conflicts of interest

There are no conflicts to declare.

## Acknowledgments

G. Z. acknowledges the support of the National Science Foundation 2012140. P.S. acknowledges the support of the Thermo Fisher Scientific fellowship.

## Notes and references

1. D. Psaltis, S. R. Quake and C. Yang, *nature*, 2006, **442**, 381-386.
2. C. Monat, P. Domachuk and B. Eggleton, *Nature photonics*, 2007, **1**, 106-114.
3. Y. Zhao, Z. S. Stratton, F. Guo, M. I. Lapsley, C. Y. Chan, S.-C. S. Lin and T. J. Huang, *Lab on a Chip*, 2013, **13**, 17-24.
4. J. Wu, G. Zheng and L. M. Lee, *Lab on a Chip*, 2012, **12**, 3566-3575.
5. Y. Zhao, D. Chen, H. Yue, J. B. French, J. Rufo, S. J. Benkovic and T. J. Huang, *Lab on a Chip*, 2013, **13**, 2183-2198.
6. A. Isozaki, J. Harmon, Y. Zhou, S. Li, Y. Nakagawa, M. Hayashi, H. Mikami, C. Lei and K. Goda, *Lab on a Chip*, 2020, **20**, 3074-3090.
7. L. Pang, H. M. Chen, L. M. Freeman and Y. Fainman, *Lab on a Chip*, 2012, **12**, 3543-3551.
8. D. V. Vezenov, B. T. Mayers, R. S. Conroy, G. M. Whitesides, P. T. Snee, Y. Chan, D. G. Nocera and M. G. Bawendi, *Journal of the American Chemical Society*, 2005, **127**, 8952-8953.
9. S. Xiong, A. Q. Liu, L. K. Chin and Y. Yang, *Lab on a Chip*, 2011, **11**, 1864-1869.
10. W. Song and D. Psaltis, *Lab on a Chip*, 2011, **11**, 2397-2402.
11. X. Cui, L. M. Lee, X. Heng, W. Zhong, P. W. Sternberg, D. Psaltis and C. Yang, *Proceedings of the National Academy of Sciences*, 2008, **105**, 10670-10675.
12. G. Zheng, S. A. Lee, S. Yang and C. Yang, *Lab on a Chip*, 2010, **10**, 3125-3129.
13. W. Bishara, H. Zhu and A. Ozcan, *Optics express*, 2010, **18**, 27499-27510.
14. V. Bianco, M. Paturzo, V. Marchesano, I. Gallotta, E. Di Schiavi and P. Ferraro, *Lab on a Chip*, 2015, **15**, 2117-2124.
15. S. A. Lee, R. Leitao, G. Zheng, S. Yang, A. Rodriguez and C. Yang, *PloS one*, 2011, **6**, e26127.
16. W. Hoppe, *Acta Crystallographica Section a-Crystal Physics Diffraction Theoretical and General Crystallography*, 1969, 495-&.
17. H. M. L. Faulkner and J. Rodenburg, *Physical review letters*, 2004, **93**, 023903.
18. M. Dierolf, A. Menzel, P. Thibault, P. Schneider, C. M. Kewish, R. Wepf, O. Bunk and F. Pfeiffer, *Nature*, 2010, **467**, 436-439.
19. J. Marrison, L. Räty, P. Marriott and P. O'toole, *Scientific reports*, 2013, **3**, 1-7.
20. A. M. Maiden, J. M. Rodenburg and M. J. Humphry, *Optics letters*, 2010, **35**, 2585-2587.
21. A. M. Maiden, M. J. Humphry, F. Zhang and J. M. Rodenburg, *JOSA A*, 2011, **28**, 604-612.
22. E. Balaur, G. A. Cadenazzi, N. Anthony, A. Spurling, E. Hanssen, J. Orian, K. A. Nugent, B. S. Parker and B. Abbey, *Nature Photonics*, 2021, **15**, 222-229.
23. D. F. Gardner, M. Tanksalvala, E. R. Shanblatt, X. Zhang, B. R. Galloway, C. L. Porter, R. Karl Jr, C. Bevis, D. E. Adams and H. C. Kapteyn, *Nature Photonics*, 2017, **11**, 259-263.
24. G. Zheng, R. Horstmeyer and C. Yang, *Nature photonics*, 2013, **7**, 739.
25. G. Zheng, C. Shen, S. Jiang, P. Song and C. Yang, *Nature Reviews Physics*, 2021, **3**, 207-223.
26. F. Pfeiffer, *Nature Photonics*, 2018, **12**, 9-17.
27. J. Rodenburg and A. Maiden, in *Springer Handbook of Microscopy*, Springer, 2019, ch. 17, pp. 819-904.
28. I. M. Vellekoop, A. Lagendijk and A. Mosk, *Nature photonics*, 2010, **4**, 320-322.
29. Y. Choi, T. D. Yang, C. Fang-Yen, P. Kang, K. J. Lee, R. R. Dasari, M. S. Feld and W. Choi, *Physical Review Letters*, 2011, **107**, 023902.
30. E. G. van Putten, D. Akbulut, J. Bertolotti, W. L. Vos, A. Lagendijk and A. Mosk, *Physical review letters*, 2011, **106**, 193905.
31. J.-H. Park, C. Park, H. Yu, J. Park, S. Han, J. Shin, S. H. Ko, K. T. Nam, Y.-H. Cho and Y. Park, *Nature photonics*, 2013, **7**, 454.
32. Y. Choi, C. Yoon, M. Kim, W. Choi and W. Choi, *IEEE Journal of Selected Topics in Quantum Electronics*, 2013, **20**, 61-73.

33. Z. Bian, S. Jiang, P. Song, H. Zhang, P. Hoveida, K. Hoshino and G. Zheng, *Journal of Physics D: Applied Physics*, 2019, **53**, 014005.

34. P. Song, S. Jiang, H. Zhang, Z. Bian, C. Guo, K. Hoshino and G. Zheng, *Optics letters*, 2019, **44**, 3645-3648.

35. S. Jiang, J. Zhu, P. Song, C. Guo, Z. Bian, R. Wang, Y. Huang, S. Wang, H. Zhang and G. Zheng, *Lab on a Chip*, 2020, **20**, 1058-1065.

36. P. Song, R. Wang, J. Zhu, T. Wang, Z. Bian, Z. Zhang, K. Hoshino, M. Murphy, S. Jiang and C. Guo, *Optics Letters*, 2020, **45**, 3486-3489.

37. M. Guizar-Sicairos and J. R. Fienup, *Optics express*, 2008, **16**, 7264-7278.

38. D. Batey, T. Edo, C. Rau, U. Wagner, Z. Pešić, T. Waigh and J. Rodenburg, *Physical Review A*, 2014, **89**, 043812.

39. W. Xu, M. Jericho, I. Meinertzhagen and H. Kreuzer, *Proceedings of the National Academy of Sciences*, 2001, **98**, 11301-11305.

40. R. C. Hardie, K. J. Barnard, J. G. Bognar, E. E. Armstrong and E. A. Watson, *Optical Engineering*, 1998, **37**, 247-260.

41. K. G. Larkin, M. A. Oldfield and H. Klemm, *Optics communications*, 1997, **139**, 99-106.

42. M. Guizar-Sicairos, S. T. Thurman and J. R. Fienup, *Optics letters*, 2008, **33**, 156-158.

43. H. Zhang, S. Jiang, J. Liao, J. Deng, J. Liu, Y. Zhang and G. Zheng, *Optics express*, 2019, **27**, 7498-7512.

44. S. A. Lee, J. Erath, G. Zheng, X. Ou, P. Willems, D. Eichinger, A. Rodriguez and C. Yang, *PloS one*, 2014, **9**, e89712.

45. K. Guo, Z. Zhang, S. Jiang, J. Liao, J. Zhong, Y. C. Eldar and G. Zheng, *Biomedical optics express*, 2018, **9**, 260-275.